\documentclass[aps,prd,reprint,superscriptaddress,notitlepage,nofootinbib]{revtex4-1}

\usepackage[utf8]{inputenc}
\usepackage[english]{babel}
\usepackage{mathtools}
\usepackage{subdepth}
\usepackage{natbib}
\usepackage[bottom]{footmisc}
\usepackage{kotex}
\usepackage{graphicx}
\usepackage{float}
\usepackage{amssymb}
\usepackage{amsmath}
\usepackage{xcolor}
\usepackage{hyperref}
\usepackage{slashed}
\usepackage{kotex}
\usepackage[shortlabels]{enumitem}
\usepackage{microtype}
\usepackage{cancel}
\usepackage{comment}
\usepackage[normalem]{ulem}
\usepackage{soul}
\usepackage{cancel}

\usepackage{booktabs}

\begin{document}

\title{Primordial Black Hole Hotspots Beyond Flat Spacetime}

\author{Doojin Kim}
\email[]{doojin.kim@usd.edu}
\affiliation{Department of Physics, University of South Dakota, Vermillion, SD 57069, USA}

\author{TaeHun Kim}
\email[]{gimthcha@kias.re.kr}
\affiliation{School of Physics, Korea Institute for Advanced Study, 85 Hoegi-ro, Dongdaemun-gu, Seoul 02455, Republic of Korea}
\affiliation{International Center for Quantum-field Measurement Systems for Studies of the Universe and Particles, High Energy Accelerator Research Organization, Oho 1-1, Tsukuba, Ibaraki 305-0801, Japan}

\author{Jong-Chul Park}
\email[]{jcpark@cnu.ac.kr}
\affiliation{Department of Physics, Chungnam National University, Daejeon, 34134, Republic of Korea}
\affiliation{Institute for Sciences of the Universe, Chungnam National University, Daejeon 34134, Republic of Korea}

\author{Jong-Hyun Yoon}
\email[]{yoonjh@cnu.ac.kr}
\affiliation{Department of Physics, Chungnam National University, Daejeon, 34134, Republic of Korea}
\affiliation{Institute of Quantum Systems, Chungnam National University, Daejeon, 34134, Republic of Korea}


\begin{abstract}
Light primordial black holes heat the surrounding plasma via Hawking radiation, forming localized hotspots whose temperature may far exceed that of the cosmological background. 
Previous studies of hotspot formation and cooling have treated the subsequent energy transport in flat spacetime, thereby neglecting the expansion of the Universe. 
We formulate the diffusion equation governing the hotspot evolution, in an expanding universe, and clarify the regime in which the formalism is valid. 
We find that hotspot formation is robust against cosmological expansion. We show that the critical distance scale, where Hubble expansion overtakes diffusion, coincides with the decoupling radius introduced in earlier work, and the temperature profile $T\propto r^{-7/11}$ essentially remains unchanged. 
However, the cooling stage is substantially modified. 
We find that the plateau temperature of a cooling hotspot initially undergoes a rapid drop and then follows $T_{\rm plt} \propto t^{-11/15}$, steeper than the flat-spacetime scaling $t^{-7/15}$.
This scaling cannot be obtained by simply redshifting the flat-spacetime solution, because expansion also suppresses diffusive transport.
As a consequence, all hotspots disappear within a finite time, as opposed to the flat-spacetime prediction of everlasting hotspots in part of the parameter space.
\end{abstract}

\maketitle

\allowdisplaybreaks

\section{Introduction}
\label{sec:intro}

Primordial black holes (PBHs) are black holes that may have formed in the early Universe from the collapse of primordial overdensities~\cite{Zeldovich:1967lct, Carr:1974nx, Carr:1975qj}, from inflation~\cite{Yokoyama:1995ex, Garcia-Bellido:1996mdl, Garcia-Bellido:2017mdw}, first-order phase transitions~\cite{Hawking:1982ga, Moss:1994iq, Khlopov:1998nm, Jung:2021mku, Hong:2020est, Kawana:2021tde, Lu:2022paj, Kawana:2022lba, Lu:2022jnp, Lu:2022yuc, Marfatia:2024cac, Liu:2021svg, Kawana:2022olo}, and others~\cite{Ruffini:1969qy, Hawking:1987bn, Cotner:2016cvr, Cotner:2017tir, Amendola:2017xhl, Cotner:2018vug, Cotner:2019ykd, Conzinu:2020cke, Flores:2020drq, DeLuca:2022bjs, Kim:2024gqp, Holst:2024ubt}. 
Once formed, they undergo evaporation via Hawking radiation, emitting high-energy particles with a characteristic Hawking temperature $T_H$ given by $T_H = M_{\rm Pl}^2/M$~\cite{Hawking:1974rv, Hawking:1975vcx, Hawking:1976de, Page:1976df, Page:1976ki, Page:1977um}, where $M_{\rm Pl}$ and $M$ denote the Planck mass and the PBH mass, respectively. 
Since the Hawking temperature far exceeds the temperature of the ambient plasma at the time of evaporation, the emitted particles deposit energy into the surrounding medium, thereby heating it. 
In particular, for PBHs lighter than $10^9 \, \text{g}$, evaporation is completed before the Big Bang nucleosynthesis, allowing the injected energy to give rise to various cosmological consequences in the early Universe while remaining consistent with observational constraints~\cite{Carr:2009jm, Carr:2020gox, Keith:2020jww}.

The cosmological impact of such evaporating PBHs has been extensively discussed in terms of global effects, such as entropy production, dark-matter emission and phenomenology, modifications of the thermal history, reheating, induced gravitational waves, and isocurvature perturbations~\cite{Acharya:2020jbv, Inomata:2020lmk, Baldes:2020nuv, Papanikolaou:2020qtd, Domenech:2020ssp, Masina:2020xhk, Cheek:2021odj, Domenech:2021wkk, Papanikolaou:2022chm, Agashe:2022phd, Jho:2022wxd, RiajulHaque:2023cqe, Gehrman:2023esa, Kim:2023ixo, Ghoshal:2023sfa,  Domenech:2024wao,  Kim:2025kgu, Franciolini:2026fdv}.
Importantly, since Hawking radiation deposits its energy \textit{locally} through scattering with particles in the surrounding plasma, the usual homogeneous treatment should be regarded as a coarse-grained description of many localized regions of elevated temperature---often called \emph{hotspots}---around individual evaporating PBH~\cite{Das:2021wei,He:2022wwy,He:2024wvt}. 
The maximum temperature attained within each hotspot can exceed that predicted by homogeneous treatment by many orders of magnitude, opening the possibility of genuinely local high-energy phenomena such as sphaleron-mediated baryogenesis, production of topological defects, dark-matter production, and catalyzed vacuum decay~\cite{Das:2021wei,He:2022wwy,Hamaide:2023ayu}.

The diffusion of energy injected by Hawking radiation into the ambient plasma was first analyzed in Ref.~\cite{Das:2021wei}, where the steady-state profile $T \propto r^{-1/3}$ was derived and the post-evaporation cooling behavior was discussed. 
The Landau-Pomeranchuk-Migdal (LPM) effect~\cite{Landau:1953um,Migdal:1956tc,Arnold:2002ja}, which suppresses the thermalization of high-energy particles in the surrounding plasma, was incorporated in Ref.~\cite{He:2022wwy}, where it was shown that the maximum hotspot temperature is independent of the initial PBH mass. 
A subsequent study~\cite{He:2024wvt} numerically confirmed these estimates by solving the Boltzmann equations governing the Hawking radiation cascade in conjunction with the diffusion equation. 
The phenomenological implications of hotspots for baryogenesis and dark-matter production were further explored in Ref.~\cite{Gunn:2024xaq}, while the role of hotspots in nucleosynthesis and the effect of modified evaporation histories were studied in Refs.~\cite{Altomonte:2025hpt,Levy:2025lyj}.
 
However, all of these analyses neglect the expansion of the Universe. 
The diffusion equation used in Refs.~\cite{Das:2021wei,He:2022wwy,He:2024wvt} was formulated under flat spacetime, thereby omitting both the Hubble dilution term $4H\rho$ and the scale-factor suppression of spatial gradients. 
Since $t \sim {H}^{-1}$ in both the radiation-dominated and matter-dominated eras, (light) PBHs evaporating during these eras have $\tau_{\rm PBH}\simeq t_{\rm ev}\sim H^{-1}(t_{\rm ev})$, where $\tau_{\rm PBH}$ and $t_{\rm ev}$ denote the PBH lifetime and the evaporation time, respectively. 
Thus, Hubble expansion can affect the hotspot evolution on a comparable time scale and is not \textit{a priori} negligible.

In this work, we incorporate the Hubble expansion into the analysis of PBH hotspot dynamics. 
We formulate the diffusion equation in an expanding background and clarify the conditions under which the diffusion approximation is valid. 
We identify the critical length scale $R_H = \sqrt{\lambda v / H}$, where $\lambda$ and $v$ denote the mean free path and particle velocity, respectively, beyond which the Hubble expansion dominates over diffusion. 
We further show that $R_H$ coincides with the decoupling radius $r_{\rm dec}$ defined in Ref.~\cite{He:2022wwy} throughout radiation domination. 
We then numerically solve the Hubble-corrected diffusion equation during the cooling phase and find that the late-time plateau temperature scales as $T_{\rm plt} \propto t^{-11/15}$, steeper than the $t^{-7/15}$ scaling obtained without the Hubble expansion~\cite{Das:2021wei}, thereby ensuring that every hotspot disappears within a finite time. 
We also verify that the envelope profile $T \propto r^{-7/11}$~\cite{He:2022wwy} is robust against cosmological corrections.
Notably, this result cannot be obtained by naively redshifting the flat-spacetime solution, since the Hubble expansion simultaneously suppresses the efficiency of diffusion.
 
The rest of this paper is organized as follows. 
In Sec.~\ref{sec:hotspot_review}, we briefly review the formation of hotspots around evaporating PBHs. 
In Sec.~\ref{sec:diffusion}, we formulate the diffusion equation in an expanding universe, establish its validity conditions, identify the critical Hubble scale, and verify the robustness of the resulting temperature profile. 
In Sec.~\ref{sec:cooling_expanding}, we solve the dimensionless cooling equation numerically. 
In Sec.~\ref{sec:hubble_effect}, we map the results to physical parameters and analyze the effect of Hubble expansion, focusing on the hotspot disappearance time. 
We summarize and discuss in Sec.~\ref{sec:conclusion}. 
We summarize the quantities defined and used in this work in Appendix~\ref{sec:notations}.
Throughout this paper, we work in natural units with $c = \hbar = k_B = 1$.

\section{Hotspot formation around evaporating PBHs}
\label{sec:hotspot_review}
 
In this section, we briefly review the formation of hotspots around evaporating PBHs, based on the discussions in Refs.~\cite{Das:2021wei,He:2022wwy,He:2024wvt}. 
This sets up the initial conditions for the cooling analysis in Sec.~\ref{sec:cooling_expanding}.
 
A PBH of mass $M$ emits Hawking radiation at a temperature $T_H = M_{\rm Pl}^2 / M$. 
The emitted high-energy particles undergo successive splittings in the ambient plasma, with the splitting rate at high energies suppressed by the LPM effect~\cite{Landau:1953um,Migdal:1956tc,Arnold:2002ja}. 
The LPM-suppressed splitting rate for a particle of energy $E$ in a plasma of temperature $T$ scales as $\Gamma_{\rm LPM} \sim \alpha^2 T \sqrt{T/E}$~\cite{He:2022wwy,Arnold:2002ja}, where $\alpha$ is the fine-structure constant of the gauge interaction responsible for the splitting.\footnote{Not necessarily restricted to gauge interactions though.} 
The emitted hard Hawking particles with $E \sim T_H \gg T$ begin interacting with the ambient plasma immediately after emission, but due to the LPM suppression, their energy is not converted efficiently into local thermal energy until a characteristic radius $r_* \sim \Gamma_{\rm LPM}^{-1}$ that is many orders of magnitude larger than the black hole horizon~\cite{He:2022wwy,He:2024wvt}. 
Therefore, the hotspot temperature should be understood as an effective, coarse-grained measure of the energy deposited by the Hawking-induced cascade, since full thermalization takes place only after the emitted hard particles propagate over $r_*$.
 
The energy injection is balanced by outward diffusion, leading to a quasi-steady effective temperature profile for the heated plasma. 
For $r < r_*$, the temperature is approximately uniform, forming a central plateau region, while for $r > r_*$, it falls as $T \propto r^{-1/3}$~\cite{Das:2021wei,He:2022wwy}. 
The plateau temperature increases as the PBH evaporates and reaches its maximum value near the final stage of evaporation~\cite{He:2022wwy},
\begin{equation}
    T_{\rm max} \simeq 0.02 \, \alpha^{19/3}\, g_*^{-4/3}\, g_{H*}^{5/6}\, M_{\rm Pl} \,,
    \label{eq:Tmax}
\end{equation}
where $g_*$ and $g_{H*}$ are the effective numbers of relativistic degrees of freedom in the ambient plasma and in the Hawking radiation, respectively. 
Remarkably, this expression is independent of the initial PBH mass, as both the energy injection rate and the thermalization radius depend on $M$ in a way that they cancel each other.

As the PBH evaporates and loses its mass, the Hawking temperature increases as $T_H \propto M^{-1}$, while the plateau temperature rises in proportion ($T_{\rm plt} \propto T_H$~\cite{He:2022wwy}). 
The accelerated LPM splitting in the hotter plasma then causes the thermalization radius to shrink as 
\begin{equation}
    r_*(M) \sim \Gamma_{\rm LPM}^{-1} \propto T_H^{-1}(M) \propto M\,,
\end{equation}
decreasing in step with the PBH mass.
As the effective source region shrinks, shells of heated plasma formed at progressively smaller radii decouple from it and subsequently cool via diffusion. 
Each decoupled shell then expands and redshifts, and the superposition of all such shells at the completion of PBH evaporation gives rise to an envelope profile $T \propto r^{-7/11}$. 
The resulting temperature profile at $t = t_{\rm ev}$, namely when the PBH has fully evaporated, serves as the initial condition for the subsequent cooling stage~\cite{He:2022wwy}:
\begin{equation}
    T(r, t_{\rm ev}) = T_{\rm max} \times
    \begin{cases}
        1 & r \leq r_0 \,, \\
        (r/r_0)^{-7/11} & r_0 < r \leq r_{\rm dec} \,,
    \end{cases}
    \label{eq:Tprofile_evap}
\end{equation}
where the plateau radius $r_0$ is
\begin{equation}
    r_0 \simeq 6 \times 10^7 \left(\frac{\alpha}{0.1}\right)^{-6} \left(\frac{g_*}{106.75}\right) \left(\frac{g_{H*}}{108}\right)^{-1} T_H^{-1}(M_*)\,, \label{eq:r0}
\end{equation}
and the decoupling radius $r_{\rm dec}$ is
\begin{eqnarray}
    r_{\rm dec} &\simeq& 3 \times 10^{-10}\,\text{sec}\, \left(\frac{\alpha}{0.1}\right)^{-8/5} \left(\frac{g_*}{106.75}\right)^{1/5} \nonumber \\
    &&\times \left(\frac{g_{H*}}{108}\right)^{-4/5} \left(\frac{T_{H,\rm ini}}{10^4\,\text{GeV}}\right)^{-11/5} \,. \label{eq:rdec}
\end{eqnarray}
Here, $M_*$ is the critical PBH mass at which the diffusion time scale becomes comparable to the evaporation time scale, while $T_{H,\rm ini}$ is the initial Hawking temperature. 
Beyond $r_{\rm dec}$, diffusion does not have sufficient time to transport the injected energy over the course of the PBH lifetime, i.e. it marks the outermost radius that can be reached by diffusion during the PBH lifetime.
 
Once the PBH has fully evaporated, the energy source no longer exists and the hotspot enters a cooling phase. 
Its subsequent evolution is governed solely by the diffusion equation, with the previously described temperature profile---consisting of a central plateau and an outer envelope---serving as the initial condition. 
It is this cooling phase that we investigate in detail in the following sections, now including the effects of the Hubble expansion.

\section{Diffusion in an expanding universe}
\label{sec:diffusion}

In this section, we formulate the diffusion process in an expanding universe by first introducing the Hubble-corrected diffusion equation. 
We then examine the conditions under which the diffusion description is valid, since it provides a collective, coarse-grained account of the random walks of individual particles. 
Finally, we confirm that the temperature profile in Eq.~\eqref{eq:Tprofile_evap} remains valid even in the presence of the Hubble expansion.

\subsection{Diffusion equation}
\label{sec:diffusion_eq}

For a relativistic plasma with energy density $\rho$, mean free path $\lambda$, and particle velocity $v$, the net energy flux is given by $\vec{J}_E = -(\lambda v / 3) \vec\nabla \rho$~\cite{reif1965}. 
Combining this relation with energy conservation $\partial_t \rho + \vec\nabla\cdot\vec{J}_E = 0$ (a.k.a. continuity equation or transport equation), one obtains the standard diffusion equation in flat spacetime:
\begin{equation}
    \frac{\partial \rho}{\partial t}=\vec{\nabla} \cdot \left( \frac{\lambda v}{3} \vec\nabla \rho \right). \label{eq:flatdiffuse}
\end{equation}
Here, the spatial gradient is done with respect to the physical coordinate.

In a Friedmann-Robertson-Walker background, however, two effects essentially modify the above flat-spacetime diffusion equation. 
First, for a relativistic species with $p = \rho/3$, the covariant conservation of the stress-energy tensor $\nabla_\mu T^{\mu\nu} = 0$ gives
\begin{equation}
    \frac{\partial \rho}{\partial t} + 4H\rho = 0
    \label{eq:rho_redshift}
\end{equation}
in the absence of diffusion, reproducing the standard scaling behavior $\rho \propto a^{-4}$. 
Second, since the physical and comoving gradients are related by $\vec\nabla_{\rm phys} = a^{-1}\vec\nabla_c$, the flux term and its divergence are respectively modified to
\begin{eqnarray}
    \vec{J}_E &=& -\frac{\lambda v}{3a}\vec\nabla_c \rho \,, \\
    \quad \vec\nabla_{\rm phys}\cdot\vec{J}_E &=& -\frac{1}{a^2}\vec\nabla_c\cdot\left(\frac{\lambda v}{3}\vec\nabla_c \rho\right) \,.
    \label{eq:flux_comoving}
\end{eqnarray}
Including the diffusion term in Eq.~\eqref{eq:rho_redshift} and using Eq.~\eqref{eq:flux_comoving}, we obtain the expanding-universe analog of the cosmic-ray diffusion equation~\cite{Berezinsky:2005fa}, now written for the energy density:
\begin{equation}
    \frac{\partial \rho}{\partial t} + 4H\rho = \frac{1}{a^2} \vec\nabla_c \cdot \left( \frac{\lambda v}{3} \vec\nabla_c \, \rho \right) \,.
    \label{eq:diffusion_full}
\end{equation}
For a plasma in thermal equilibrium with $g_*$ relativistic degrees of freedom, $\rho = g_* \pi^2 T^4 / 30$, while the mean free path can be dimensionally estimated as 
\begin{equation}
    \lambda \sim \frac{1}{n\sigma}\sim \frac{1}{T^3 (\alpha^2/T^2)}=\frac{1}{\alpha^2 T}\,,
\end{equation}
where we assume that the relevant scattering process is governed only by dimensionless coupling.

Compared to the flat-spacetime diffusion equation in Eq.~\eqref{eq:flatdiffuse} adopted in Refs.~\cite{Das:2021wei,He:2022wwy,He:2024wvt}, Eq.~\eqref{eq:diffusion_full} contains two expansion-induced modifications: the Hubble dilution term $4H\rho$ and the factor $1/a^2$ in front of the spatial derivatives. 
The former accounts for cooling due to cosmological redshift, while the latter suppresses the efficiency of diffusion as the Universe expands.

\subsection{Validity conditions}
\label{sec:validity}

Having formulated the Hubble-corrected diffusion equation, we now examine the conditions under which the diffusion approximation is valid. 
In particular, the energy flux equation $\vec{J}_E = -(\lambda v/3)\vec\nabla\rho$ relies on the assumption that the medium is approximately homogeneous over one mean free path. This requires two independent conditions.

The first condition ensures that the mean free path is well defined, i.e., $\lambda$ does not change appreciably during a single scattering event. A particle travels a distance $\lambda$ in a time interval $\Delta t = \lambda/v$, during which the mean free path changes by 
\begin{equation}
    \Delta\lambda = \frac{d\lambda}{dt}\Delta t=\frac{d\lambda}{dt}\frac{\lambda}{v}\,.
\end{equation}
Now, requiring $|\Delta\lambda| \ll \lambda$ leads to
\begin{equation}
    \left| \frac{d\lambda}{dt} \right| \ll v \,.
    \label{eq:cond1}
\end{equation}
Since $\lambda \sim 1/(\alpha^2 T)$ and temperature $T$ generally varies in space and time, this condition becomes
\begin{equation}
    \left| \frac{\partial T}{\partial t} + \vec{v} \cdot \vec\nabla T \right| \ll \alpha^2 T^2 v \,.
    \label{eq:cond1_expanded}
\end{equation}
In the stationary limit, i.e. $\partial_t T = 0$, this condition reduces to $|\nabla T|/T^2 \ll \alpha^2$, which coincides with the local thermal equilibrium condition given in Ref.~\cite{He:2024wvt}. 
In this context, Eq.~\eqref{eq:cond1} is more general because it also captures temporal variations of the medium. 
Since Eq.~\eqref{eq:cond1_expanded} depends on the solution $T(r,t)$, it must be checked \emph{a posteriori}. 
To this end, we define a dimensionless quantity
\begin{equation}
\mathcal{V}(r,t) \equiv \frac{1}{\alpha^2 T^2}\left|\frac{\partial T}{\partial t}+v\frac{\partial T}{\partial r}\right| ,
\end{equation}
and confirm numerically that $\mathcal{V} \lesssim \mathcal{O}(10^{-2})$ throughout the hotspot over the entire time range considered in this work.

The other condition ensures that particles undergo many scatterings within a Hubble time, so that their transport can be described as diffusion:
\begin{equation}
\frac{v}{\lambda} \gg H \,.
\label{eq:cond2}
\end{equation}
Equation~\eqref{eq:cond2} compares the scattering rate with the expansion rate, whereas Eq.~\eqref{eq:cond1} constrains the \emph{variation} of $\lambda$ over a single scattering event. 
Both conditions must be satisfied independently.

\subsection{Critical scale for the Hubble expansion}
\label{sec:critical_scale}

The effect of the Hubble expansion on the hotspot evolution might be seemingly negligible, since a hotspot occupies a region much smaller than the Hubble radius. 
Nevertheless, its evolution can still be sensitive to cosmic expansion if its dynamical time scale is cosmological.
This motivates us to identify the critical length scale below which diffusion can overcome cosmological expansion, by comparing the Hubble time $H^{-1}$ with the time required for diffusion to reach a given distance.

A particle undergoing a random walk with mean free path $\lambda$ and velocity $v$ diffuses over a distance $r_{\rm diff}(t) \sim \sqrt{\lambda v t}$ in a given time interval $t$. 
Equivalently, the time required for diffusion to redistribute energy across a given scale $r$ is 
\begin{equation} 
    t_{\rm diff}(r) \sim \frac{r^2}{\lambda v}\,.
\end{equation}
Diffusion then dominates the Hubble dilution only if the diffusion time scale is less than the Hubble time, i.e., $t_{\rm diff}(r) \lesssim H^{-1}$, which defines a critical length scale\footnote{Of course, the same expression can also be obtained directly by replacing $t$ in the diffusion length scale $\sqrt{\lambda v t}$ with the Hubble time $H^{-1}$.}
\begin{equation}
    R_H = \sqrt{\frac{\lambda v}{H}}\,.
\label{eq:RH}
\end{equation}
On scales $r \ll R_H$, diffusion can maintain the temperature profile against cosmological expansion, whereas for $r \gtrsim R_H$, the Hubble dilution becomes the main driver of the energy-density evolution. 
Since the diffusion validity condition~\eqref{eq:cond2} requires $\lambda \ll v/H$, we immediately find that $R_H \ll v/H$. 
The critical length scale is therefore always well below the causal horizon.

During the radiation-dominated era, $H = 1/(2t)$, hence $R_H \sim \sqrt{\lambda v t}$, which coincide with the diffusion length scale $r_{\rm diff}(t)$. Evaluated at the evaporation time $t = t_{\rm ev}$, this gives
\begin{equation}
    R_H(t_{\rm ev}) \sim \sqrt{\lambda v\, t_{\rm ev}} = r_{\rm dec} \,,
    \label{eq:RH_rdec}
\end{equation}
where $r_{\rm dec}$ is identified as the decoupling radius defined in Ref.~\cite{He:2022wwy}. 
More generally, the same identification holds at any time; $R_H(t) = r_{\rm diff}(t)$. 
In other words, the diffusion length always marks the scale at which the Hubble expansion begins to overtake diffusive transport. 
This in turn provides a physical interpretation of $r_{\rm dec}$ as the boundary between the diffusion-dominated and Hubble-dominated regimes.

\subsection{Robustness of the $r^{-7/11}$ initial profile}
\label{sec:robustness}

So far, we have established that the diffusion-based description of hotspot formation remains valid in the presence of the Hubble expansion. 
We now further verify that cosmological redshift results in only a negligible correction to the hotspot temperature profile at the completion moment of PBH evaporation, so that the $T \propto r^{-7/11}$~\cite{He:2022wwy} and the central plateau characterized by Eq.~\eqref{eq:Tmax} remain intact.

As discussed in Sec.~\ref{sec:hotspot_review}, the envelope is a parametric curve traced by shells of heated plasma that decouple from PBH as it evaporates. 
If a given shell decouples when the PBH mass has decreased to a fraction $m$ of its initial mass, $M' = m  M_{\rm ini}$, then the decoupling time is given by $t_{\rm ev}(1 - m^3)$.
By the time $t_{\rm ev}$, when the PBH completely evaporates, cosmological expansion modifies the temperature and physical radius of a given shell relative to their flat-spacetime values by factors
\begin{align}
    \frac{T(m)}{T_{\rm flat}(m)} &= (1-m^3)^{1/2}\,, \label{eq:THub} \\
    \frac{r(m)}{r_{\rm flat}(m)} &= (1-m^3)^{-1/2}\,, \label{eq:rHub}
\end{align}
where the subscript ``flat'' denotes the corresponding quantities obtained in the absence of the Hubble expansion.

Since $t_{\rm ev} \propto M^3$, evaporation accelerates strongly near the end of the PBH lifetime. 
As a result, the inner shells $(m \ll 1)$ decouple only shortly before $t_{\rm ev}$ and therefore experience negligible redshift. More generally, expansion-induced corrections to $T(m)$ and $r(m)$ remain smooth and modest over the relevant range of $m$: they are about 6\% at $m=0.5$ and fall below the percent level for $m \lesssim 0.1$. 
Because these corrections do not introduce any sharp $m$-dependence, the parametric relation between temperature and radius is only weakly distorted, so the envelope slope $\dfrac{d\log T}{d\log r}=-\dfrac{7}{11}$ remains a very good approximation over the dynamically relevant range.

\section{Cooling of hotspot in an expanding universe}
\label{sec:cooling_expanding}

Having established that the formation of hotspot is essentially unaffected by the Hubble expansion, we now turn to the cooling stage that begins immediately after the PBH evaporation. 
In this section, we solve the diffusion equation~\eqref{eq:diffusion_full} including the Hubble expansion, and then compare the result with that without the Hubble expansion in the next section. 

In the relativistic regime, we use the scalings $\lambda \propto 1/T$ and $\rho \propto T^4$. 
To make the temperature dependence of the mean free path explicit, we define $\tilde{\lambda} \equiv T\lambda \sim 1/\alpha^2$. 
Substituting $\rho = g_* \pi^2 T^4/30$ into Eq.~\eqref{eq:diffusion_full} and assuming spherical symmetry, Eq.~\eqref{eq:diffusion_full} can be rewritten as an evolution equation for the radial temperature profile $T(r_c, t)$: 
\begin{equation}
    \frac{\partial T}{\partial t} = - H T + \frac{\tilde{\lambda} v}{3 a^2 T^2} \left[T \left(\frac{\partial^2 T}{\partial r_c^2} + \frac{2}{r_c} \frac{\partial T}{\partial r_c} \right) + 2\left(\frac{\partial T}{\partial r_c} \right)^2 \right], \label{eq:Tradial}
\end{equation}
where $r_c$ denotes the comoving radius. Assuming radiation domination, the Hubble rate scales as $H = 1/2t \propto a^{-2}$. Since the Hubble redshift causes the temperature to scale as $T \propto 1/a \propto 1/\sqrt{t}$, it is convenient to define a redshift-corrected temperature
\begin{equation}
    U(r_c, t) = \sqrt{\frac{t}{t_{\rm ev}}} \, T(r_c, t) \label{eq:U}
\end{equation}
so that $U = T$ at $t = t_{\rm ev}$ and $U$ remains constant during the subsequent cooling stage, $t>t_{\rm ev}$, if the redshift is the only effect at work. 
Substituting Eq.~\eqref{eq:U} into Eq.~\eqref{eq:Tradial}, we then find the evolution equation for $U$:
\begin{eqnarray}
    \frac{\partial U}{\partial t} &=& \frac{\tilde{\lambda}v}{3 a_{\rm ev}^2 \sqrt{t/t_{\rm ev}} \, U^2} \nonumber \\
    &&\times \left[U \left(\frac{\partial^2 U}{\partial r_c^2}+\frac{2}{r_c} \frac{\partial U}{\partial r_c} \right) + 2\left(\frac{\partial U}{\partial r_c} \right)^2 \right], \label{eq:Udot}
\end{eqnarray}
where the redshift term has been absorbed into the definition of $U$ on the left-hand side. 
Here, $a_{\rm ev}$ is the scale factor at $t_{\rm ev}$ with $a_{\rm ev}=a\sqrt{t_{\rm ev}/t}$, and we choose the normalization $a_{\rm ev} = 1$.

Remarkably, Eq.~\eqref{eq:Udot} can be further simplified by defining dimensionless variables: 
\begin{eqnarray}
    u &\equiv& \frac{U}{T_{\rm max}} = \sqrt{\frac{t}{t_{\rm ev}}} \frac{T}{T_{\rm max}} \,, \label{eq:u} \\
    \tau &\equiv& \frac{2\tilde{\lambda} v}{3 T_{\rm max} a_{\rm ev}^2 r_0^2} \sqrt{t_{\rm ev}} \left(\sqrt{t} - \sqrt{t_{\rm ev}} \right) \,, \label{eq:tau} \\
    x &\equiv& \frac{r_c}{r_0} \,, \label{eq:x} 
\end{eqnarray}
which define a dimensionless redshift-corrected temperature, a dimensionless time variable, and a dimensionless comoving radius, respectively; we calibrate $\tau = 0$ at $t = t_{\rm ev}$. 
In terms of $u$, $\tau$, and $x$, Eq.~\eqref{eq:Udot} becomes
\begin{equation}
    \frac{\partial u}{\partial \tau} = \frac{1}{u} \left(\frac{\partial^2 u}{\partial x^2} + \frac{2}{x} \frac{\partial u}{\partial x} \right) + \frac{2}{u^2} \left( \frac{\partial u}{\partial x} \right)^2\,, \label{eq:uEvolEq}
\end{equation}
which is \textit{independent of any explicit physical parameters}; all model dependence enters only through the conversion back to physical units.

We recall that Eq.~\eqref{eq:Tprofile_evap} provides the initial temperature profile at $t=t_{\rm ev}$. Since the subsequent evolution is formulated in terms of the comoving radius $r_c$, we rewrite this profile using $r=a_{\rm ev}r_c$. With the chosen normalization $a_{\rm ev}=1$, this simply amounts to identifying $r$ with $r_c$. The temperature profile is now expressed as
\begin{equation}
    u(x, \tau = 0) = 
    \begin{cases}
        1 & \text{for} \quad x \leq 1\\
        x^{-7/11} & \text{for} \quad x > 1
	\end{cases} \,, \label{eq:ucoolinginitial}
\end{equation}
which is also independent of any explicit physical parameters. 
Therefore, Eq.~\eqref{eq:uEvolEq} needs to be solved numerically only once, again since its evolution is universal and does not explicitly depend on the physical parameters. 
All parameter dependence enters only through the transformations in Eqs.~\eqref{eq:u}--\eqref{eq:x}.

\begin{figure}[t]
    \centering
    \includegraphics[width=0.98\linewidth]{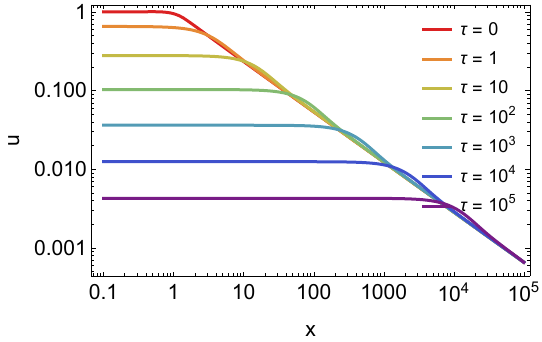}
    \caption{Evolution of the radial profiles of the dimensionless redshift-corrected temperature $u(x)$ at representative values of the dimensionless time $\tau$.}
    \label{fig:uprofile}
\end{figure}

In Fig.~\ref{fig:uprofile}, we present the numerical evolution of the radial dimensionless redshift-corrected temperature profiles $u(x)$ for representative values of the dimensionless time $\tau$, starting from the initial condition specified in Eq.~\eqref{eq:ucoolinginitial} at $\tau = 0$. 
We find that the profiles retain their qualitative shape, consisting of a central plateau region and an outer envelope that continues to obey the same scaling $u \propto x^{-7/11}$. 
The only quantity that evolves with time is the plateau height, which we denote by $u_{\rm plt}$. 
The plateau radius increases accordingly, being determined by the junction with the envelope. 
Therefore, the entire hotspot temperature profile during the cooling stage can be determined once $u_{\rm plt} (\tau)$ is specified.

Rather than solving Eq.~\eqref{eq:diffusion_full} directly to extract the functional form, we can take a simpler route. 
As seen in Fig.~\ref{fig:uprofile}, a point in the outer profile begins to deviate from its initial envelope behavior, $u=x^{-7/11}$, precisely when it is reached by the expanding plateau boundary. 
This observation allows us to infer the plateau radius as a function of time, and hence $u_{\rm plt}(\tau)$ through Eq.~\eqref{eq:ucoolinginitial}.
We therefore begin with substituting $u = x^{-7/11}$ into Eq.~\eqref{eq:uEvolEq} and find $\partial u / \partial \tau = 70/(121x^2)$ in the envelope region. 
Assuming that this initial rate remains approximately valid for short times, the accumulated change is given by $\Delta u = 70/(121x^2) \tau$. 
We estimate the plateau boundary, choosing the location where this correction becomes comparable to the initial envelope value, i.e., $\Delta u \sim u(\tau=0)= x^{-7/11}$.
This gives $x = [(70/121) \tau]^{11/15}$, providing a rough estimate of the plateau boundary.
Finally, through Eq.~\eqref{eq:ucoolinginitial}, we then obtain the plateau height at $\tau$
\begin{equation}
    u_{\rm plt}(\tau) \sim \left( \frac{70}{121} \tau \right)^{-7/15} \,. \label{eq:uplaanal}
\end{equation}

\begin{figure}[t]
    \centering
    \includegraphics[width=0.98\linewidth]{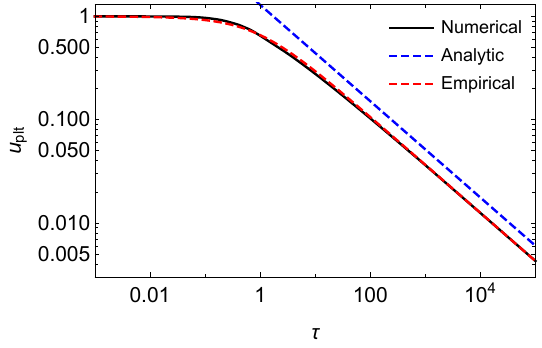}
    \caption{Evolution of the plateau height $u_{\rm plt}$ as a function of $\tau$. 
    Black: numerical result. 
    Blue dashed: analytic expression in Eq.~\eqref{eq:uplaanal}. 
    Red dashed: empirical expression in Eq.~\eqref{eq:uplaemp}.}
    \label{fig:uplateaufitting}
\end{figure}

However, this expression incorrectly predicts $u_{\rm plt} \rightarrow \infty$ as $\tau\to0$, whereas the correct initial value is $u_{\rm plt} = 1$ at $\tau = 0$. 
This comparison is presented in Fig.~\ref{fig:uplateaufitting}, where the numerical result is plotted in black and Eq.~\eqref{eq:uplaanal} is shown by the blue dashed curve. 
In addition to the failure near $\tau \rightarrow 0$, the analytic estimate also exhibits a mild $\mathcal{O}(1)$ offset in the asymptotic regime. 
Motivated by the scaling behavior in Eq.~\eqref{eq:uplaanal}, we find the following simple empirical fit that reproduces the numerical result with a good enough accuracy:
\begin{equation}
    u_{\rm plt} \simeq (1.2 \tau + 0.25 \sqrt{\tau} + 1 ) ^{-7/15} \,. \label{eq:uplaemp}
\end{equation}
We plot Eq.~\eqref{eq:uplaemp} by the red dashed curve in Fig.~\ref{fig:uplateaufitting}, and use it in the rest of our analysis. 
We emphasize again that Figs.~\ref{fig:uprofile}, \ref{fig:uplateaufitting}, and Eq.~\eqref{eq:uplaemp} are universal in the sense that they do not depend explicitly on any combination of physical parameters.

The same calculation can be performed in the absence of the Hubble expansion. 
Starting from Eq.~\eqref{eq:diffusion_full}, we set $H = 0$ and $a = 1$, and replace the comoving radius $r_c$ by the physical radius $r$. 
Repeating the same rescaling procedure, we again arrive at the same equations, Eqs.~\eqref{eq:uEvolEq} and \eqref{eq:ucoolinginitial}; the only difference lies in the definitions of the dimensionless quantities in Eqs.~\eqref{eq:u}--\eqref{eq:x}, which are now given by 
\begin{eqnarray}
    u &\equiv& \frac{T}{T_{\rm max}} \,, \label{eq:unoHub} \\
    \tau &\equiv& \frac{\tilde{\lambda} v}{3 T_{\rm max} r_0^2} (t - t_{\rm ev}) \,, \label{eq:taunoHub} \\
    x &\equiv& \frac{r}{r_0} \,. \label{eq:xnoHub} 
\end{eqnarray}
Therefore, the dimensionless solutions shown in Figs.~\ref{fig:uprofile} and \ref{fig:uplateaufitting} remain unchanged. 
Only the conversion back to physical parameters differs.

\section{Effect of the Hubble expansion in hotspot cooling}
\label{sec:hubble_effect}

We are now in a position to consider the physical parameters and discuss the effect of the Hubble expansion in the cooling stage and possible phenomenology. 

\begin{figure}[t]
    \centering
    \includegraphics[width=0.98\linewidth]{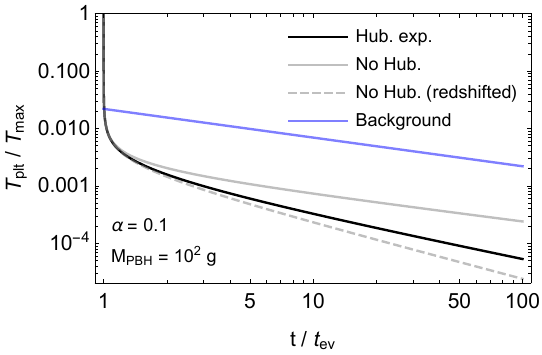} \\
    \includegraphics[width=0.98\linewidth]{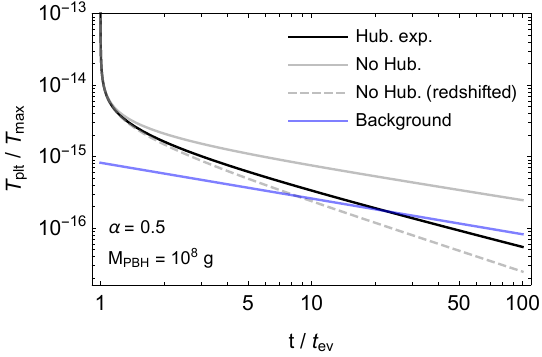} 
    \caption{Evolution of the $T_{\max}$-normalized plateau temperature as a function of time (black), compared with the case which neglects the Hubble expansion (gray), a naive redshift correction to it (gray dashed), and the background temperature of the Universe (blue).}
    \label{fig:ToverT0examples}
\end{figure}

\begin{figure}[t]
    \centering
    \includegraphics[width=0.98\linewidth]{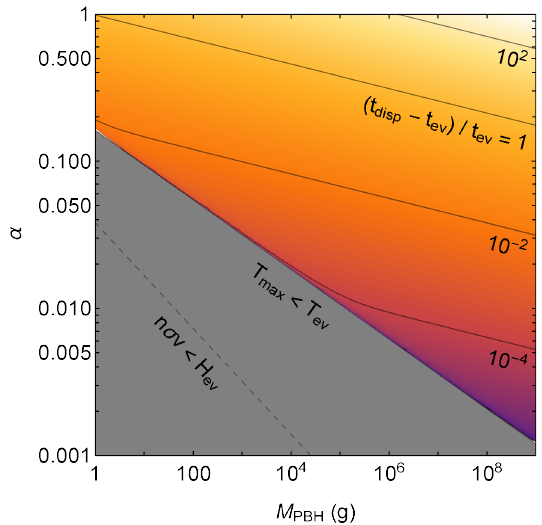} \\
    \includegraphics[width=0.98\linewidth]{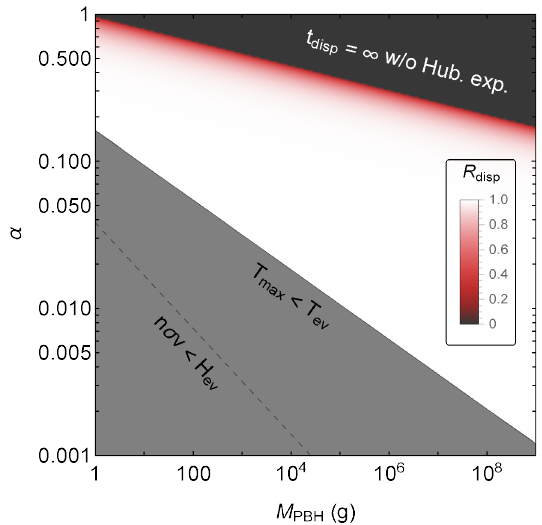}
    \caption{Contours of the hotspot disappearance time $t_{\rm disp}$ in the parameter space of $M_{\rm PBH}$ and $\alpha$. 
     Upper: Ratio of the time elapsed between PBH evaporation and hotspot disappearance, $(t_{\rm disp}-t_{\rm ev})$, to the cosmic time at evaporation, $t_{\rm ev}$. 
    Lower: Ratio of $t_{\rm disp}$ to the value obtained in the absence of the Hubble expansion, i.e., $R_{\rm disp}$ defined in Eq.~\eqref{eq:Rdisp}. 
    }
    \label{fig:tdisp}
\end{figure}

Figure~\ref{fig:ToverT0examples} shows example plots of the post-evaporation evolution of the $T_{\max}$-normalized plateau temperature $T_{\rm plt}/T_{\max}$ as a function of the physical time $t$, for two different sets of coupling and PBH mass parameter choices. 
The temperature initially decreases very rapidly as cooling begins and then transitions to a milder power-law decline. 
This early drop can span many orders of magnitude. 
This behavior is not surprising because the horizontal axis is normalized by $t_{\rm ev}$, i.e., the cosmic time at PBH evaporation, which is typically much longer than the diffusion time scale. 
As a result, even a very small time interval $t/t_{\rm ev}$ can correspond to a significant period for diffusion. 
Moreover, this hierarchy becomes even larger for heavier PBHs, since $t_{\rm ev}$ increases with $M_{\rm PBH}$ (compare the behaviors in the upper and lower panels; also see Fig.~\ref{fig:tdisp} for a complete parameter dependence).

Once the time further elapses and reaches $t - t_{\rm ev} \sim \mathcal{O}(0.1) t_{\rm ev}$, the plateau temperature enters a power-law cooling regime. 
This is the stage where the Hubble expansion governs the cooling behavior because its effect becomes significant (or at least appreciable) only on cosmological time scales. 
Using Eqs.~\eqref{eq:u}, \eqref{eq:tau}, and \eqref{eq:uplaemp}, we find that the $t$ dependence of the plateau temperature in the long-time scale is given by 
\begin{equation}
    T_{\rm plt} \propto t^{-11/15}\,. \label{eq:TplaOurs}
\end{equation}
By contrast, for the case neglecting the Hubble expansion, a similar calculation with Eqs.~\eqref{eq:unoHub}, \eqref{eq:taunoHub}, and \eqref{eq:uplaemp} gives $T_{\rm plt} \propto t^{-7/15}$~\cite{Das:2021wei}. 
A naive redshifting of the latter result would instead yield $T_{\rm plt} \propto t^{-29/30}$, which still differs from our result in Eq.~\eqref{eq:TplaOurs}. 
These three behaviors are shown by the black, gray, and gray dashed curves in Fig.~\ref{fig:ToverT0examples}. 
The difference arises because the Hubble expansion not only redshifts the temperature but also weakens diffusion through the $1/a^2$ factor multiplied to the diffusion operator in Eq.~\eqref{eq:diffusion_full}.

Based on these observations, we briefly comment on the hotspot cooling behavior investigated in Ref.~\cite{Das:2021wei}, where the scaling behavior of $T_{\rm plt} \propto t^{-7/15}$ was analytically derived. 
Our result improves their findings in two aspects. 
First, the plateau temperature is expected to undergo a substantial and rapid drop before the power-law scaling dominates. 
This can be important because in cases such as the upper panel in Fig.~\ref{fig:ToverT0examples}, the hotspot disappears during this initial rapid-cooling period, which we will discuss in detail shortly. 
Second, in the power-law cooling regime, the Hubble expansion modifies the scaling exponent from $-7/15$ to $-11/15$, which is an effect that cannot be reproduced by a naive redshifting of the result in Ref.~\cite{Das:2021wei}.

We further examine how the Hubble expansion affects the cooling by considering the hotspot disappearance time, $t_{\rm disp}$. 
We define $t_{\rm disp}$ as the time when the plateau temperature falls below the background temperature of the Universe. 
Since the background temperature is $T_{\rm ev}$ at $t_{\rm ev}$, cosmological redshift gives $T_{\rm bkg} = T_{\rm ev} \times (t/t_{\rm ev})^{-1/2}$. 
This is shown by the blue lines in Fig.~\ref{fig:ToverT0examples}. 
If the hotspot disappears during the rapid cooling stage (upper panel), the Hubble expansion makes an insignificant impact on $t_{\rm disp}$. 
However, if it survives into the power-law regime (lower panel), the effect of the Hubble expansion can be substantial. 
Without Hubble expansion, the plateau temperature follows the scaling behavior $T_{\rm plt} \propto t^{-7/15}$, a slower fall-off than the standard redshift of the background temperature. 
This implies that for some parameter choices, the hotspot would survive indefinitely. 
However, once the Hubble expansion is taken into account, the plateau temperature decreases faster than the background temperature at late times, ensuring that every hotspot disappears within a finite time.

To quantify this behavior, Fig.~\ref{fig:tdisp} shows contours of $t_{\rm disp}$ in the $(M_{\rm PBH}, \alpha)$ parameter space. 
We first exclude the gray shaded region, where our diffusion-based treatment breaks down. 
The condition $T_{\rm max} < T_{\rm ev}$ indicates that the maximum hotspot temperature predicted by the diffusion calculation is below the background temperature of the Universe at evaporation, so a well-defined hotspot is not formed within our framework. 
We, however, remark that this should not be interpreted as the non-existence of the hotspot. 
Since the Hawking temperature continues to increase toward the final stage of evaporation, some localized energy deposition may still occur, but such a regime cannot be captured by the present diffusion description and is reserved for future work. 
In addition, $n \sigma v = v/\lambda < H$ means that the emitted particles do not scatter frequently enough to thermalize within a Hubble time, also signaling the breakdown of our diffusion-based calculation.\footnote{$n\sigma v$ is temperature-dependent, and we consider its value at the plateau since it is the region where the diffusion is effective; the envelope follows $T \propto t^{-1/2}$ ($u = \text{constant}$) during the cooling, which has no difference from the redshift.}

At the moment of PBH evaporation, the valid region of the parameter space where our treatment is valid is characterized by $T_{\rm max} > T_{\rm ev}$ and $n\sigma v > H_{\rm ev}$, with the former providing the more stringent condition. 
Indeed, in the plateau region, one has $n \sigma v \sim \alpha^2 \, T_{\rm plt}$ while the background temperature satisfies $T_{\rm bkg} \sim \sqrt{M_{\rm Pl} H} \gg H$. 
Therefore, once the condition $n\sigma v > H_{\rm ev}$ is satisfied at evaporation, it remains satisfied until the hotspot disappears, namely until $T_{\rm plt} = T_{\rm bkg}$.

In the remaining parameter space, the upper panel shows the ratio of the post-evaporation disappearance time, i.e., $t_{\rm disp} - t_{\rm ev}$, to the evaporation time itself $t_{\rm ev}$. 
We find that over the majority of the parameter space, $t_{\rm disp} - t_{\rm ev} \ll t_{\rm ev}$, corresponding to the behavior shown in the upper panel of Fig.~\ref{fig:ToverT0examples}. 
However, in the upper-right corner of the parameter space, $t_{\rm disp} - t_{\rm ev} \gtrsim t_{\rm ev}$, corresponding to the lower panel of Fig.~\ref{fig:ToverT0examples}, where the cooling persists to the power-law stage. 

In the lower panel, we display the ratio of the disappearance time obtained in our analysis $t_{\rm disp}$ to the corresponding value calculated without the Hubble expansion, denoted by $t_{{\rm disp},\cancel{H}}$: 
\begin{equation}
    R_{\rm disp} \equiv \frac{t_{\rm disp}}{t_{{\rm disp},\cancel{H}}}  \,. \label{eq:Rdisp}
\end{equation}
The notable difference emerges in the region of $t_{\rm disp} - t_{\rm ev} \gtrsim t_{\rm ev}$ where the cooling stage extends to the power-law regime; in the absence of the Hubble expansion, this regime would predict everlasting hotspots. 
For $t_{\rm disp} - t_{\rm ev} \ll t_{\rm ev}$, the cooling is complete, and the hotspot disappears during the initial rapid-drop phase, leaving no considerable difference between the cases with and without the Hubble expansion.

Several phenomenological applications may be sensitive to these findings. 
For example, if a symmetry is restored or particle production is reactivated inside the hotspot because its temperature exceeds the background temperature, both the disappearance time $t_{\rm disp}$ and the time dependence of the hotspot temperature can affect the total particle production. 
This effect can be particularly important in regions of the parameter space where $t_{\rm disp} - t_{\rm ev} \gtrsim t_{\rm ev}$, for which the Hubble expansion can significantly modify the predicted evolution. 
Moreover, if hotspots give rise to observable signatures, their presence, lifetime, and disappearance time may directly affect the resulting observables. 
We leave a detailed investigation of these phenomenological implications to future work.

\section{Summary and Discussion}
\label{sec:conclusion}

In this work, we studied the dynamics of hotspots around evaporating PBHs in an expanding universe. 
Although most Hawking radiation is emitted during the late stage of evaporation, the cosmological time at evaporation satisfies $t \sim H^{-1}$, so the Hubble expansion must be taken into account for proper treatment.

We analyzed the effect of the Hubble expansion on both the formation and cooling of the hotspot. 
To this end, we first formulated the diffusion equation in an expanding background as expressed in Eq.~\eqref{eq:diffusion_full}, where the Hubble dilution term appears as $4H\rho$, and the spatial gradients in the diffusion operator are suppressed by $1/a^2$ in comoving coordinates. 

We then showed that hotspot formation is unaffected by the expansion of the Universe. 
The critical scale $R_H = \sqrt{\lambda v / H}$, beyond which the Hubble expansion begins to dominate over diffusion, coincides with the decoupling radius $r_{\rm dec}$ identified in Ref.~\cite{He:2022wwy}. 
This result follows directly from $H \sim 1/t$: the maximum distance accessible to diffusion is precisely the distance beyond which the Hubble expansion becomes relevant. Moreover, since most of the black hole’s energy is emitted near the end of the evaporation process, the temperature profile does not undergo a significant redshift.

However, we found that the cooling stage can be substantially affected. 
We solved the diffusion equation with the Hubble expansion directly by introducing dimensionless variables. 
Since both the equation and the initial condition are independent of the underlying physical parameters, our numerical results and the empirical formula in Eq.~\eqref{eq:uplaemp} are applicable to arbitrary parameter choices. 
Upon converting the results back to physical variables, we identified two important features of the cooling behavior. 
First, the plateau temperature exhibits a pronounced rapid drop before settling into a power-law decrease; for some parameter choices, the hotspot may vanish already during this stage. 
Second, in the power-law regime, we found $T_{\rm plt} \propto t^{-11/15}$, which differs from the no-expansion result, $T_{\rm plt} \propto t^{-7/15}$. 
This change in the exponent reflects the combined effects of Hubble dilution and the $1/a^2$ suppression of the diffusion operator, and thus cannot be captured by a naive manual redshift. 

Such a difference can significantly affect the hotspot disappearance time. 
Over most of the parameter space, hotspots disappear during the initial rapid-drop phase, leaving little opportunity for the Hubble expansion to modify the outcome. 
However, for heavy PBHs with large couplings, hotspots survive into the subsequent power-law regime. 
In this case, the flat-spacetime result would predict that the hotspots persist indefinitely, whereas the Hubble expansion guarantees that all hotspots disappear within a finite time. 
This change in the cooling time can, in turn, lead to differences in possible phenomenology, such as particle production occurring locally within the hotspots.

Since these evaporating PBHs cannot constitute dark matter, a realistic cosmology is expected to include a particle dark sector. 
If that sector contains a dark gauge interaction, analogous dark hotspots may arise through thermalization. 
For example, if the dark sector consists of a massless dark photon and a massive dark-charged particle $X$ with mass $m_X$ and coupling $\alpha_D$, Eq.~\eqref{eq:uEvolEq} and its solutions would apply in essentially the same form, with the replacements $\alpha \to \alpha_D$ and $g_* \to g_{*,D}$, where $g_{*,D}$ denotes the number of dark relativistic degrees of freedom. 
Depending on the value of $m_X$, these hotspots may enter the non-relativistic regime before disappearing, in which case the $X$ population becomes Boltzmann suppressed, and the diffusion description will eventually break down. 
A detailed study of such dark hotspots is left for future work.

\begin{acknowledgments}
The authors thank Minxi He for helpful discussions.
This work was supported by the National Research Foundation of Korea (NRF) grants funded by the Ministry of Science and ICT (RS-2024-00356960 and RS-2025-00559197) and by the Global - Learning \& Academic research Institution for Master's · PhD students, and Postdocs (G-LAMP) Program of the NRF grant funded by the Ministry of Education (RS-2025-25442707). 
T.H.K. is supported by KIAS Individual Grant PG095202 at Korea Institute for Advanced Study.
\end{acknowledgments}

\clearpage
\appendix

\onecolumngrid
\section{Notations and Expressions of Variables}
\label{sec:notations}

\begin{table}[h]
\centering
\caption{Notations and expressions adopted in this work. } 
\label{tab:symbol}
\begin{tabular}{  c  p{12.8cm}  p{3.5cm}}
\toprule
 Notation & Meaning & Expression \\ 
\midrule
 $t$ & Cosmic time & -- \\
 \addlinespace[0.2em]
 $t_{\rm ev}$ & $t$ at PBH evaporation & -- \\
 \addlinespace[0.2em]
 $t_{\rm disp}$ & $t$ at PBH hotspot disappearance & -- \\
 \addlinespace[0.2em]
 $t_{{\rm disp},\cancel{H}}$ & $t_{\rm disp}$ obtained in the absence of the Hubble expansion & -- \\
 \addlinespace[0.3em]
 $t_{\rm diff}(r)$ & Time required for diffusion to redistribute energy across a distance $r$ & -- \\
 \addlinespace[0.2em]
 $M$ & PBH mass & -- \\
 \addlinespace[0.2em]
 $M_{\rm ini}$ & Initial PBH mass & -- \\
 \addlinespace[0.2em]
 $M_*$ & PBH mass at which the diffusion and evaporation time scales become comparable & -- \\
 \addlinespace[0.2em]
 $T_H$ & Hawking temperature & $T_H = M_{\rm Pl}^2/M$ \\
 \addlinespace[0.2em]
 $T_{H,\rm ini}$ & Initial Hawking temperature & -- \\
 \addlinespace[0.2em]
 $T_{\rm ev}$ & Plasma temperature at PBH evaporation & -- \\
 \addlinespace[0.2em]
 $T_{\rm bkg}$ & Background plasma temperature at $t > t_{\rm ev}$ & $T_{\rm bkg} = T_{\rm ev} \times (t/t_{\rm ev})^{-1/2}$ \\
 \addlinespace[0.2em]
 $T_{\rm plt}$ & Plateau temperature & -- \\
 \addlinespace[0.2em]
 $T_{\rm max}$ & Maximum plateau temperature, attained at the moment of complete PBH evaporation & -- \\
  \addlinespace[0.2em]
 $T_{\rm flat}$ & Temperature obtained in the absence of the Hubble expansion  & -- \\
 \addlinespace[0.2em]
 $r_*$ & Thermalization radius of Hawking radiation & $r_* \sim \Gamma_{\rm LPM}^{-1}$ \\
 \addlinespace[0.2em]
 $r_0$ & Plateau radius at the moment of complete PBH evaporation & -- \\
 \addlinespace[0.2em]
 $r_{\rm dec}$ & Decoupling radius; the outermost radius reachable by diffusion during the PBH lifetime & -- \\
 \addlinespace[0.2em]
 $r_{\rm diff}(t)$ & Diffusion length at time $t$ & $r_{\rm diff} \sim \sqrt{\lambda v t}$ \\
   \addlinespace[0.2em]
 $r_{\rm flat}$ & Radius obtained in the absence of the Hubble expansion & -- \\
 \addlinespace[0.2em]
 $R_H$ & Critical length scale at which the Hubble expansion overtakes diffusion & $R_H = \sqrt{\lambda v / H}$ \\
 \addlinespace[0.2em]
 $\lambda$ & Mean free path of plasma particles & $\lambda \sim 1/(\alpha^2 T)$ \\
 \addlinespace[0.2em]
 $\tilde{\lambda}$ & Rescaled mean free path with temperature factored out & $\tilde{\lambda} \equiv T \lambda \sim 1/\alpha^2$ \\
 \addlinespace[0.2em]
 $r_c$ & Comoving radius & -- \\
 \addlinespace[0.2em]
 $a$ & Scale factor & -- \\
 \addlinespace[0.2em]
 $a_{\rm ev}$ & $a$ at PBH evaporation & -- \\
 \addlinespace[0.2em]
 $U$ & Redshift-corrected temperature & $U = T \times (t/t_{\rm ev})^{1/2}$ \\
 \addlinespace[0.2em]
 $U_{\rm plt}$ & $U$ at plateau & -- \\
 \addlinespace[0.2em]
 $u$ & Dimensionless redshift-corrected temperature & $u \equiv U/T_{\rm max}$ \\
 \addlinespace[0.2em]
 $u_{\rm plt}$ & $u$ at plateau & -- \\
 \addlinespace[0.2em]
 $\tau$ & Dimensionless time variable for the cooling stage; $\tau = 0$ at $t = t_{\rm ev}$ & See Eq.~\eqref{eq:tau}. \\
 \addlinespace[0.2em]
 $x$ & Dimensionless comoving radius & $x \equiv r_c/r_0$ \\
 \addlinespace[0.2em]
 $R_{\rm disp}$ & Ratio of $t_{\rm disp}$ to $t_{{\rm disp},\cancel{H}}$ & 
 $R_{\rm disp} \equiv t_{\rm disp} / t_{{\rm disp},\cancel{H}}$ 
 \\
 \addlinespace[0.2em]
\bottomrule
\end{tabular} 
\end{table}

\twocolumngrid

\bibliography{references}

@article{Papanikolaou:2020qtd,
    author = "Papanikolaou, Theodoros and Vennin, Vincent and Langlois, David",
    title = "{Gravitational waves from a universe filled with primordial black holes}",
    eprint = "2010.11573",
    archivePrefix = "arXiv",
    primaryClass = "astro-ph.CO",
    doi = "10.1088/1475-7516/2021/03/053",
    journal = "JCAP",
    volume = "03",
    pages = "053",
    year = "2021"
}

@article{Cotner:2017tir,
    author = "Cotner, Eric and Kusenko, Alexander",
    title = "{Primordial black holes from scalar field evolution in the early universe}",
    eprint = "1706.09003",
    archivePrefix = "arXiv",
    primaryClass = "astro-ph.CO",
    doi = "10.1103/PhysRevD.96.103002",
    journal = "Phys. Rev. D",
    volume = "96",
    number = "10",
    pages = "103002",
    year = "2017"
}

@article{Cotner:2018vug,
    author = "Cotner, Eric and Kusenko, Alexander and Takhistov, Volodymyr",
    title = "{Primordial Black Holes from Inflaton Fragmentation into Oscillons}",
    eprint = "1801.03321",
    archivePrefix = "arXiv",
    primaryClass = "astro-ph.CO",
    reportNumber = "IPMU18-0008",
    doi = "10.1103/PhysRevD.98.083513",
    journal = "Phys. Rev. D",
    volume = "98",
    number = "8",
    pages = "083513",
    year = "2018"
}

@article{Cotner:2019ykd,
    author = "Cotner, Eric and Kusenko, Alexander and Sasaki, Misao and Takhistov, Volodymyr",
    title = "{Analytic Description of Primordial Black Hole Formation from Scalar Field Fragmentation}",
    eprint = "1907.10613",
    archivePrefix = "arXiv",
    primaryClass = "astro-ph.CO",
    reportNumber = "IPMU19-0063, YITP-19-31",
    doi = "10.1088/1475-7516/2019/10/077",
    journal = "JCAP",
    volume = "10",
    pages = "077",
    year = "2019"
}

@article{Flores:2020drq,
    author = "Flores, Marcos M. and Kusenko, Alexander",
    title = "{Primordial Black Holes from Long-Range Scalar Forces and Scalar Radiative Cooling}",
    eprint = "2008.12456",
    archivePrefix = "arXiv",
    primaryClass = "astro-ph.CO",
    reportNumber = "IPMU20-0092",
    doi = "10.1103/PhysRevLett.126.041101",
    journal = "Phys. Rev. Lett.",
    volume = "126",
    number = "4",
    pages = "041101",
    year = "2021"
}

@article{Moss:1994iq,
    author = "Moss, I. G.",
    title = "{Singularity formation from colliding bubbles}",
    doi = "10.1103/PhysRevD.50.676",
    journal = "Phys. Rev. D",
    volume = "50",
    pages = "676--681",
    year = "1994"
}

@article{Hawking:1982ga,
    author = "Hawking, S. W. and Moss, I. G. and Stewart, J. M.",
    title = "{Bubble Collisions in the Very Early Universe}",
    reportNumber = "Print-82-0180 (CAMBRIDGE)",
    doi = "10.1103/PhysRevD.26.2681",
    journal = "Phys. Rev. D",
    volume = "26",
    pages = "2681",
    year = "1982"
}

@article{Liu:2021svg,
    author = "Liu, Jing and Bian, Ligong and Cai, Rong-Gen and Guo, Zong-Kuan and Wang, Shao-Jiang",
    title = "{Primordial black hole production during first-order phase transitions}",
    eprint = "2106.05637",
    archivePrefix = "arXiv",
    primaryClass = "astro-ph.CO",
    doi = "10.1103/PhysRevD.105.L021303",
    journal = "Phys. Rev. D",
    volume = "105",
    number = "2",
    pages = "L021303",
    year = "2022"
}

@article{Lu:2022paj,
    author = "Lu, Philip and Kawana, Kiyoharu and Xie, Ke-Pan",
    title = "{Old phase remnants in first-order phase transitions}",
    eprint = "2202.03439",
    archivePrefix = "arXiv",
    primaryClass = "astro-ph.CO",
    doi = "10.1103/PhysRevD.105.123503",
    journal = "Phys. Rev. D",
    volume = "105",
    number = "12",
    pages = "123503",
    year = "2022"
}

@article{Jung:2021mku,
    author = "Jung, Tae Hyun and Okui, Takemichi",
    title = "{Primordial black holes from bubble collisions during a first-order phase transition}",
    eprint = "2110.04271",
    archivePrefix = "arXiv",
    primaryClass = "hep-ph",
    reportNumber = "KEK-TH-2350",
    month = "10",
    year = "2021"
}

@article{Hong:2020est,
    author = "Hong, Jeong-Pyong and Jung, Sunghoon and Xie, Ke-Pan",
    title = "{Fermi-ball dark matter from a first-order phase transition}",
    eprint = "2008.04430",
    archivePrefix = "arXiv",
    primaryClass = "hep-ph",
    doi = "10.1103/PhysRevD.102.075028",
    journal = "Phys. Rev. D",
    volume = "102",
    number = "7",
    pages = "075028",
    year = "2020"
}

@article{Kawana:2021tde,
    author = "Kawana, Kiyoharu and Xie, Ke-Pan",
    title = "{Primordial black holes from a cosmic phase transition: The collapse of Fermi-balls}",
    eprint = "2106.00111",
    archivePrefix = "arXiv",
    primaryClass = "astro-ph.CO",
    doi = "10.1016/j.physletb.2021.136791",
    journal = "Phys. Lett. B",
    volume = "824",
    pages = "136791",
    year = "2022"
}

@article{Kawana:2022lba,
    author = "Kawana, Kiyoharu and Lu, Philip and Xie, Ke-Pan",
    title = "{First-order phase transition and fate of false vacuum remnants}",
    eprint = "2206.09923",
    archivePrefix = "arXiv",
    primaryClass = "astro-ph.CO",
    doi = "10.1088/1475-7516/2022/10/030",
    journal = "JCAP",
    volume = "10",
    pages = "030",
    year = "2022"
}

@article{Carr:1974nx,
    author = "Carr, Bernard J. and Hawking, S. W.",
    title = "{Black holes in the early Universe}",
    doi = "10.1093/mnras/168.2.399",
    journal = "Mon. Not. Roy. Astron. Soc.",
    volume = "168",
    pages = "399--415",
    year = "1974"
}

@article{Carr:1975qj,
    author = "Carr, Bernard J.",
    title = "{The Primordial black hole mass spectrum}",
    doi = "10.1086/153853",
    journal = "Astrophys. J.",
    volume = "201",
    pages = "1--19",
    year = "1975"
}

@article{Lu:2022yuc,
    author = "Lu, Philip and Takhistov, Volodymyr and Fuller, George M.",
    title = "{Signatures of a High Temperature QCD Transition in the Early Universe}",
    eprint = "2212.00156",
    archivePrefix = "arXiv",
    primaryClass = "astro-ph.CO",
    reportNumber = "IPMU22-0064, KEK-QUP-2022-0017, KEK-TH-2476, KEK-Cosmo-0303",
    doi = "10.1103/PhysRevLett.130.221002",
    journal = "Phys. Rev. Lett.",
    volume = "130",
    number = "22",
    pages = "221002",
    year = "2023"
}

@article{Lu:2022jnp,
    author = "Lu, Philip and Kawana, Kiyoharu and Kusenko, Alexander",
    title = "{Late-forming primordial black holes: Beyond the CMB era}",
    eprint = "2210.16462",
    archivePrefix = "arXiv",
    primaryClass = "astro-ph.CO",
    doi = "10.1103/PhysRevD.107.103037",
    journal = "Phys. Rev. D",
    volume = "107",
    number = "10",
    pages = "103037",
    year = "2023"
}

@article{Carr:2020gox,
    author = "Carr, Bernard and Kohri, Kazunori and Sendouda, Yuuiti and Yokoyama, Jun'ichi",
    title = "{Constraints on primordial black holes}",
    eprint = "2002.12778",
    archivePrefix = "arXiv",
    primaryClass = "astro-ph.CO",
    reportNumber = "RESCEU-03/20; KEK-Cosmo-249; KEK-TH-2199; IPMU20-0024",
    doi = "10.1088/1361-6633/ac1e31",
    journal = "Rept. Prog. Phys.",
    volume = "84",
    number = "11",
    pages = "116902",
    year = "2021"
}

@article{Carr:2009jm,
    author = "Carr, B. J. and Kohri, Kazunori and Sendouda, Yuuiti and Yokoyama, Jun'ichi",
    title = "{New cosmological constraints on primordial black holes}",
    eprint = "0912.5297",
    archivePrefix = "arXiv",
    primaryClass = "astro-ph.CO",
    reportNumber = "RESCEU-31-09, TU-852, YITP-09-112",
    doi = "10.1103/PhysRevD.81.104019",
    journal = "Phys. Rev. D",
    volume = "81",
    pages = "104019",
    year = "2010"
}

@article{Domenech:2020ssp,
    author = "Dom\`enech, Guillem and Lin, Chunshan and Sasaki, Misao",
    title = "{Gravitational wave constraints on the primordial black hole dominated early universe}",
    eprint = "2012.08151",
    archivePrefix = "arXiv",
    primaryClass = "gr-qc",
    reportNumber = "YITP-20-156",
    doi = "10.1088/1475-7516/2021/11/E01",
    journal = "JCAP",
    volume = "04",
    pages = "062",
    year = "2021",
    note = "[Erratum: JCAP 11, E01 (2021)]"
}

@article{Inomata:2020lmk,
    author = "Inomata, Keisuke and Kawasaki, Masahiro and Mukaida, Kyohei and Terada, Takahiro and Yanagida, Tsutomu T.",
    title = "{Gravitational Wave Production right after a Primordial Black Hole Evaporation}",
    eprint = "2003.10455",
    archivePrefix = "arXiv",
    primaryClass = "astro-ph.CO",
    reportNumber = "IPMU 20-0029, DESY 20-042, DESY-20-042, CTPU-PTC-20-05",
    doi = "10.1103/PhysRevD.101.123533",
    journal = "Phys. Rev. D",
    volume = "101",
    number = "12",
    pages = "123533",
    year = "2020"
}

@article{Domenech:2021wkk,
    author = "Dom\`enech, Guillem and Takhistov, Volodymyr and Sasaki, Misao",
    title = "{Exploring evaporating primordial black holes with gravitational waves}",
    eprint = "2105.06816",
    archivePrefix = "arXiv",
    primaryClass = "astro-ph.CO",
    reportNumber = "IPMU21-0028, YITP-21-44",
    doi = "10.1016/j.physletb.2021.136722",
    journal = "Phys. Lett. B",
    volume = "823",
    pages = "136722",
    year = "2021"
}

@article{Marfatia:2024cac,
    author = "Marfatia, Danny and Tseng, Po-Yan and Yeh, Yu-Min",
    title = "{Phenomenology of bubble size distributions in a first-order phase transition}",
    eprint = "2407.15419",
    archivePrefix = "arXiv",
    primaryClass = "hep-ph",
    month = "7",
    year = "2024"
}

@article{Conzinu:2020cke,
    author = "Conzinu, P. and Gasperini, M. and Marozzi, G.",
    title = "{Primordial Black Holes from Pre-Big Bang inflation}",
    eprint = "2004.08111",
    archivePrefix = "arXiv",
    primaryClass = "gr-qc",
    reportNumber = "BA-TH/800-20",
    doi = "10.1088/1475-7516/2020/08/031",
    journal = "JCAP",
    volume = "08",
    pages = "031",
    year = "2020"
}

@article{Hawking:1987bn,
    author = "Hawking, S. W.",
    title = "{Black Holes From Cosmic Strings}",
    reportNumber = "Print-88-0310 (CAMBRIDGE)",
    doi = "10.1016/0370-2693(89)90206-2",
    journal = "Phys. Lett. B",
    volume = "231",
    pages = "237--239",
    year = "1989"
}

@article{Ruffini:1969qy,
    author = "Ruffini, Remo and Bonazzola, Silvano",
    title = "{Systems of selfgravitating particles in general relativity and the concept of an equation of state}",
    doi = "10.1103/PhysRev.187.1767",
    journal = "Phys. Rev.",
    volume = "187",
    pages = "1767--1783",
    year = "1969"
}

@article{Garcia-Bellido:2017mdw,
    author = "Garcia-Bellido, Juan and Ruiz Morales, Ester",
    title = "{Primordial black holes from single field models of inflation}",
    eprint = "1702.03901",
    archivePrefix = "arXiv",
    primaryClass = "astro-ph.CO",
    reportNumber = "IFT-UAM-CSIC-17-007, CERN-TH-2017-196",
    doi = "10.1016/j.dark.2017.09.007",
    journal = "Phys. Dark Univ.",
    volume = "18",
    pages = "47--54",
    year = "2017"
}

@article{Yokoyama:1995ex,
    author = "Yokoyama, Junichi",
    title = "{Formation of MACHO primordial black holes in inflationary cosmology}",
    eprint = "astro-ph/9509027",
    archivePrefix = "arXiv",
    reportNumber = "YITP-U-95-26",
    journal = "Astron. Astrophys.",
    volume = "318",
    pages = "673",
    year = "1997"
}

@article{Garcia-Bellido:1996mdl,
    author = "Garcia-Bellido, Juan and Linde, Andrei D. and Wands, David",
    title = "{Density perturbations and black hole formation in hybrid inflation}",
    eprint = "astro-ph/9605094",
    archivePrefix = "arXiv",
    reportNumber = "SU-ITP-96-20, SUSSEX-AST-96-5-1",
    doi = "10.1103/PhysRevD.54.6040",
    journal = "Phys. Rev. D",
    volume = "54",
    pages = "6040--6058",
    year = "1996"
}

@article{Hawking:1974rv,
    author = "Hawking, S. W.",
    title = "{Black hole explosions}",
    doi = "10.1038/248030a0",
    journal = "Nature",
    volume = "248",
    pages = "30--31",
    year = "1974"
}

@article{Hawking:1976de,
    author = "Hawking, S. W.",
    title = "{Black Holes and Thermodynamics}",
    doi = "10.1103/PhysRevD.13.191",
    journal = "Phys. Rev. D",
    volume = "13",
    pages = "191--197",
    year = "1976"
}

@article{Page:1976df,
    author = "Page, Don N.",
    title = "{Particle Emission Rates from a Black Hole: Massless Particles from an Uncharged, Nonrotating Hole}",
    doi = "10.1103/PhysRevD.13.198",
    journal = "Phys. Rev. D",
    volume = "13",
    pages = "198--206",
    year = "1976"
}

@article{Acharya:2020jbv,
    author = "Acharya, Sandeep Kumar and Khatri, Rishi",
    title = "{CMB and BBN constraints on evaporating primordial black holes revisited}",
    eprint = "2002.00898",
    archivePrefix = "arXiv",
    primaryClass = "astro-ph.CO",
    doi = "10.1088/1475-7516/2020/06/018",
    journal = "JCAP",
    volume = "06",
    pages = "018",
    year = "2020"
}

@article{DeLuca:2022bjs,
    author = "De Luca, Valerio and Franciolini, Gabriele and Riotto, Antonio",
    title = "{Heavy Primordial Black Holes from Strongly Clustered Light Black Holes}",
    eprint = "2210.14171",
    archivePrefix = "arXiv",
    primaryClass = "astro-ph.CO",
    doi = "10.1103/PhysRevLett.130.171401",
    journal = "Phys. Rev. Lett.",
    volume = "130",
    number = "17",
    pages = "171401",
    year = "2023"
}

@article{Kawana:2022olo,
    author = "Kawana, Kiyoharu and Kim, TaeHun and Lu, Philip",
    title = "{PBH formation from overdensities in delayed vacuum transitions}",
    eprint = "2212.14037",
    archivePrefix = "arXiv",
    primaryClass = "astro-ph.CO",
    doi = "10.1103/PhysRevD.108.103531",
    journal = "Phys. Rev. D",
    volume = "108",
    number = "10",
    pages = "103531",
    year = "2023"
}

@article{Papanikolaou:2022chm,
    author = "Papanikolaou, Theodoros",
    title = "{Gravitational waves induced from primordial black hole fluctuations: the~effect of an extended mass function}",
    eprint = "2207.11041",
    archivePrefix = "arXiv",
    primaryClass = "astro-ph.CO",
    doi = "10.1088/1475-7516/2022/10/089",
    journal = "JCAP",
    volume = "10",
    pages = "089",
    year = "2022"
}

@article{Domenech:2024wao,
    author = {Dom\`enech, Guillem and Tr\"ankle, Jan},
    title = "{From formation to evaporation: Induced gravitational wave probes of the primordial black hole reheating scenario}",
    eprint = "2409.12125",
    archivePrefix = "arXiv",
    primaryClass = "gr-qc",
    month = "9",
    year = "2024"
}

@article{Khlopov:1998nm,
    author = "Khlopov, M. Yu. and Konoplich, R. V. and Rubin, S. G. and Sakharov, A. S.",
    title = "{Formation of black holes in first order phase transitions}",
    eprint = "hep-ph/9807343",
    archivePrefix = "arXiv",
    reportNumber = "ROME1-1203-1998",
    month = "7",
    year = "1998"
}

@article{Das:2021wei,
    author = "Das, Saurav and Hook, Anson",
    title = "{Black hole production of monopoles in the early universe}",
    eprint = "2109.00039",
    archivePrefix = "arXiv",
    primaryClass = "hep-ph",
    doi = "10.1007/JHEP12(2021)145",
    journal = "JHEP",
    volume = "12",
    pages = "145",
    year = "2021"
}

@article{He:2022wwy,
    author = "He, Minxi and Kohri, Kazunori and Mukaida, Kyohei and Yamada, Masaki",
    title = "{Formation of hot spots around small primordial black holes}",
    eprint = "2210.06238",
    archivePrefix = "arXiv",
    primaryClass = "hep-ph",
    reportNumber = "KEK-TH-2458, KEK-Cosmo-0298, KEK-QUP-2022-0020, TU-1171",
    doi = "10.1088/1475-7516/2023/01/027",
    journal = "JCAP",
    volume = "01",
    pages = "027",
    year = "2023"
}

@book{reif1965,
  author = {Reif, F.},
  title = {Fundamentals of Statistical and Thermal Physics},
  publisher = {McGraw-Hill},
  year = {1965},
  address = {New York}
}

@article{He:2024wvt,
    author = "He, Minxi and Kohri, Kazunori and Mukaida, Kyohei and Yamada, Masaki",
    title = "{Thermalization and hotspot formation around small primordial black holes}",
    eprint = "2407.15926",
    archivePrefix = "arXiv",
    primaryClass = "hep-ph",
    reportNumber = "KEK-TH-2639, EK-TH-2639, TU-1238, CTPU-PTC-24-22, KEK-Cosmo-0351,
  KEK-QUP-2024-0018",
    doi = "10.1088/1475-7516/2024/10/080",
    journal = "JCAP",
    volume = "10",
    pages = "080",
    year = "2024"
}

@article{Hawking:1975vcx,
    author = "Hawking, S. W.",
    editor = "Gibbons, G. W. and Hawking, S. W.",
    title = "{Particle Creation by Black Holes}",
    doi = "10.1007/BF02345020",
    journal = "Commun. Math. Phys.",
    volume = "43",
    pages = "199--220",
    year = "1975",
    note = "[Erratum: Commun.Math.Phys. 46, 206 (1976)]"
}

@article{Hamaide:2023ayu,
    author = "Hamaide, Louis and Heurtier, Lucien and Hu, Shi-Qian and Cheek, Andrew",
    title = "{Primordial black holes are true vacuum nurseries}",
    eprint = "2311.01869",
    archivePrefix = "arXiv",
    primaryClass = "hep-ph",
    doi = "10.1016/j.physletb.2024.138895",
    journal = "Phys. Lett. B",
    volume = "856",
    pages = "138895",
    year = "2024"
}

@article{Gunn:2024xaq,
    author = "Gunn, Jacob and Heurtier, Lucien and Perez-Gonzalez, Yuber F. and Turner, Jessica",
    title = "{Primordial black hole hot spots and out-of-equilibrium dynamics}",
    eprint = "2409.02173",
    archivePrefix = "arXiv",
    primaryClass = "hep-ph",
    reportNumber = "KCL-PH-TH-2024-48, IPPP/24/58",
    doi = "10.1088/1475-7516/2025/02/040",
    journal = "JCAP",
    volume = "02",
    pages = "040",
    year = "2025"
}

@article{Landau:1953um,
    author = "Landau, L. D. and Pomeranchuk, I.",
    title = "{Limits of applicability of the theory of bremsstrahlung electrons and pair production at high-energies}",
    journal = "Dokl. Akad. Nauk Ser. Fiz.",
    volume = "92",
    pages = "535--536",
    year = "1953"
}

@article{Migdal:1956tc,
    author = "Migdal, A. B.",
    title = "{Bremsstrahlung and Pair Production at High Energies in Condensed Media}",
    doi = "10.1103/PhysRev.103.1811",
    journal = "Phys. Rev.",
    volume = "103",
    pages = "1811--1820",
    year = "1956"
}

@article{Arnold:2002ja,
    author = "Arnold, Peter Brockway and Moore, Guy D. and Yaffe, Laurence G.",
    title = "{Photon and gluon emission in relativistic plasmas}",
    eprint = "hep-ph/0204343",
    archivePrefix = "arXiv",
    reportNumber = "UW-PT-02-06",
    doi = "10.1088/1126-6708/2002/06/030",
    journal = "JHEP",
    volume = "06",
    pages = "030",
    year = "2002"
}

@article{Altomonte:2025hpt,
    author = "Altomonte, Clelia and Fairbairn, Malcolm and Heurtier, Lucien",
    title = "{Primordial black hole hot spots and nucleosynthesis}",
    eprint = "2501.05531",
    archivePrefix = "arXiv",
    primaryClass = "astro-ph.CO",
    doi = "10.1103/mvxw-vny6",
    journal = "Phys. Rev. D",
    volume = "112",
    number = "12",
    pages = "123057",
    year = "2025"
}

@article{Levy:2025lyj,
    author = "Levy, Nathaniel and Heurtier, Lucien",
    title = "{Effect of the memory burden on primordial black hole hot spots}",
    eprint = "2511.17329",
    archivePrefix = "arXiv",
    primaryClass = "astro-ph.CO",
    doi = "10.1103/zhrp-yj6l",
    journal = "Phys. Rev. D",
    volume = "113",
    number = "4",
    pages = "043037",
    year = "2026"
}

@article{Zeldovich:1967lct,
    author = "Zel'dovich, Ya. B. and Novikov, I. D.",
    title = "{The Hypothesis of Cores Retarded during Expansion and the Hot Cosmological Model}",
    journal = "Sov. Astron.",
    volume = "10",
    pages = "602",
    year = "1967"
}

@article{Page:1976ki,
    author = "Page, Don N.",
    title = "{Particle Emission Rates from a Black Hole. 2. Massless Particles from a Rotating Hole}",
    doi = "10.1103/PhysRevD.14.3260",
    journal = "Phys. Rev. D",
    volume = "14",
    pages = "3260--3273",
    year = "1976"
}

@article{Page:1977um,
    author = "Page, Don N.",
    title = "{Particle Emission Rates from a Black Hole. 3. Charged Leptons from a Nonrotating Hole}",
    doi = "10.1103/PhysRevD.16.2402",
    journal = "Phys. Rev. D",
    volume = "16",
    pages = "2402--2411",
    year = "1977"
}

@article{RiajulHaque:2023cqe,
    author = "Riajul Haque, Md and Kpatcha, Essodjolo and Maity, Debaprasad and Mambrini, Yann",
    title = "{Primordial black hole reheating}",
    eprint = "2305.10518",
    archivePrefix = "arXiv",
    primaryClass = "hep-ph",
    doi = "10.1103/PhysRevD.108.063523",
    journal = "Phys. Rev. D",
    volume = "108",
    number = "6",
    pages = "063523",
    year = "2023"
}

@article{Berezinsky:2005fa,
    author = "Berezinsky, Veniamin and Gazizov, A. Z.",
    title = "{Diffusion of cosmic rays in expanding universe}",
    eprint = "astro-ph/0512090",
    archivePrefix = "arXiv",
    doi = "10.1086/502626",
    journal = "Astrophys. J.",
    volume = "643",
    pages = "8--13",
    year = "2006"
}

@article{Amendola:2017xhl,
    author = "Amendola, Luca and Rubio, Javier and Wetterich, Christof",
    title = "{Primordial black holes from fifth forces}",
    eprint = "1711.09915",
    archivePrefix = "arXiv",
    primaryClass = "astro-ph.CO",
    doi = "10.1103/PhysRevD.97.081302",
    journal = "Phys. Rev. D",
    volume = "97",
    number = "8",
    pages = "081302",
    year = "2018"
}

@article{Cotner:2016cvr,
    author = "Cotner, Eric and Kusenko, Alexander",
    title = "{Primordial black holes from supersymmetry in the early universe}",
    eprint = "1612.02529",
    archivePrefix = "arXiv",
    primaryClass = "astro-ph.CO",
    reportNumber = "IPMU16-0192",
    doi = "10.1103/PhysRevLett.119.031103",
    journal = "Phys. Rev. Lett.",
    volume = "119",
    number = "3",
    pages = "031103",
    year = "2017"
}

@article{Kim:2024gqp,
    author = "Kim, Taehun and Lu, Philip",
    title = "{Primordial black hole reformation in the early Universe}",
    eprint = "2411.07469",
    archivePrefix = "arXiv",
    primaryClass = "astro-ph.CO",
    doi = "10.1016/j.physletb.2025.139488",
    journal = "Phys. Lett. B",
    volume = "865",
    pages = "139488",
    year = "2025"
}

@article{Holst:2024ubt,
    author = "Holst, Ian and Krnjaic, Gordan and Xiao, Huangyu",
    title = "{Clustering and runaway merging in a primordial black hole dominated universe}",
    eprint = "2412.01890",
    archivePrefix = "arXiv",
    primaryClass = "astro-ph.CO",
    reportNumber = "FERMILAB-PUB-24-0888-PPD-T",
    doi = "10.1103/296w-v86n",
    journal = "Phys. Rev. D",
    volume = "112",
    number = "8",
    pages = "083527",
    year = "2025"
}

@article{Keith:2020jww,
    author = "Keith, Celeste and Hooper, Dan and Blinov, Nikita and McDermott, Samuel D.",
    title = "{Constraints on Primordial Black Holes From Big Bang Nucleosynthesis Revisited}",
    eprint = "2006.03608",
    archivePrefix = "arXiv",
    primaryClass = "astro-ph.CO",
    reportNumber = "FERMILAB-PUB-20-224-A",
    doi = "10.1103/PhysRevD.102.103512",
    journal = "Phys. Rev. D",
    volume = "102",
    number = "10",
    pages = "103512",
    year = "2020"
}

@article{Kim:2025kgu,
    author = "Kim, TaeHun and Gong, Jinn-Ouk and Jeong, Donghui and Jung, Dong-Won and Kim, Yeong Gyun and Lee, Kang Young",
    title = "{Planck isocurvature constraint on primordial black holes lighter than a kiloton}",
    eprint = "2503.14581",
    archivePrefix = "arXiv",
    primaryClass = "astro-ph.CO",
    reportNumber = "APCTP-Pre2025-005",
    month = "3",
    year = "2025"
}

@article{Baldes:2020nuv,
    author = "Baldes, Iason and Decant, Quentin and Hooper, Deanna C. and Lopez-Honorez, Laura",
    title = "{Non-Cold Dark Matter from Primordial Black Hole Evaporation}",
    eprint = "2004.14773",
    archivePrefix = "arXiv",
    primaryClass = "astro-ph.CO",
    reportNumber = "ULB-TH/20-05",
    doi = "10.1088/1475-7516/2020/08/045",
    journal = "JCAP",
    volume = "08",
    pages = "045",
    year = "2020"
}

@article{Masina:2020xhk,
    author = "Masina, Isabella",
    title = "{Dark matter and dark radiation from evaporating primordial black holes}",
    eprint = "2004.04740",
    archivePrefix = "arXiv",
    primaryClass = "hep-ph",
    doi = "10.1140/epjp/s13360-020-00564-9",
    journal = "Eur. Phys. J. Plus",
    volume = "135",
    number = "7",
    pages = "552",
    year = "2020"
}

@article{Ghoshal:2023sfa,
    author = "Ghoshal, Anish and Gouttenoire, Yann and Heurtier, Lucien and Simakachorn, Peera",
    title = "{Primordial black hole archaeology with gravitational waves from cosmic strings}",
    eprint = "2304.04793",
    archivePrefix = "arXiv",
    primaryClass = "hep-ph",
    reportNumber = "IPPP/23/13",
    doi = "10.1007/JHEP08(2023)196",
    journal = "JHEP",
    volume = "08",
    pages = "196",
    year = "2023"
}

@article{Gehrman:2023esa,
    author = "Gehrman, Thomas C. and Shams Es Haghi, Barmak and Sinha, Kuver and Xu, Tao",
    title = "{The primordial black holes that disappeared: connections to dark matter and MHz-GHz gravitational Waves}",
    eprint = "2304.09194",
    archivePrefix = "arXiv",
    primaryClass = "hep-ph",
    reportNumber = "UTWI-10-2023",
    doi = "10.1088/1475-7516/2023/10/001",
    journal = "JCAP",
    volume = "10",
    pages = "001",
    year = "2023"
}

@article{Kim:2023ixo,
    author = "Kim, TaeHun and Lu, Philip and Marfatia, Danny and Takhistov, Volodymyr",
    title = "{Regurgitated dark matter}",
    eprint = "2309.05703",
    archivePrefix = "arXiv",
    primaryClass = "hep-ph",
    reportNumber = "KEK-QUP-2023-0019, KEK-TH-2550, KEK-Cosmo-0321, IPMU23-0029",
    doi = "10.1103/PhysRevD.110.L051702",
    journal = "Phys. Rev. D",
    volume = "110",
    number = "5",
    pages = "L051702",
    year = "2024"
}

@article{Franciolini:2026fdv,
    author = "Franciolini, G. and Racco, D.",
    title = "{Isocurvature Constraints on Dark Matter from Evaporated Primordial Black Holes}",
    eprint = "2603.02322",
    archivePrefix = "arXiv",
    primaryClass = "astro-ph.CO",
    month = "3",
    year = "2026"
}

@article{Cheek:2021odj,
    author = "Cheek, Andrew and Heurtier, Lucien and Perez-Gonzalez, Yuber F. and Turner, Jessica",
    title = "{Primordial black hole evaporation and dark matter production. I. Solely Hawking radiation}",
    eprint = "2107.00013",
    archivePrefix = "arXiv",
    primaryClass = "hep-ph",
    reportNumber = "FERMILAB-PUB-21-304-T, NUHEP-TH/21-06, CP3-21-41, IPPP/21/02",
    doi = "10.1103/PhysRevD.105.015022",
    journal = "Phys. Rev. D",
    volume = "105",
    number = "1",
    pages = "015022",
    year = "2022"
}

@article{Jho:2022wxd,
    author = "Jho, Yongsoo and Kim, Tae-Geun and Park, Jong-Chul and Park, Seong Chan and Park, Yeji",
    title = "{Primordial Black Holes as a Factory of Axions: Extragalactic Photons from Axions}",
    eprint = "2212.11977",
    archivePrefix = "arXiv",
    primaryClass = "hep-ph",
    doi = "10.1093/ptep/ptag011",
    journal = "PTEP",
    volume = "2026",
    number = "2",
    pages = "023B07",
    year = "2026"
}

@article{Agashe:2022phd,
    author = "Agashe, Kaustubh and Chang, Jae Hyeok and Clark, Steven J. and Dutta, Bhaskar and Tsai, Yuhsin and Xu, Tao",
    title = "{Detecting axionlike particles with primordial black holes}",
    eprint = "2212.11980",
    archivePrefix = "arXiv",
    primaryClass = "hep-ph",
    reportNumber = "UMD-PP-022-12, MI-HET-792",
    doi = "10.1103/PhysRevD.108.023014",
    journal = "Phys. Rev. D",
    volume = "108",
    number = "2",
    pages = "023014",
    year = "2023"
}

\end{document}